\documentclass[aps,prl,twocolumn,10pt,floatfix,superscriptaddress]{revtex4-2}
\usepackage{amssymb}
\usepackage{amsmath}
\usepackage{graphicx}% Include figure files
\usepackage{dcolumn}% Align table columns on decimal point
\usepackage{bm}% bold math

\usepackage{comment}
\usepackage[dvipsnames]{xcolor}
\usepackage{hyperref}% add hypertext capabilities
\hypersetup{
   colorlinks,%
   citecolor=blue,%
   filecolor=blue,%
   linkcolor=blue,%
   urlcolor=blue
}

\begin{document}

\title{Spatial structure and magnetism of a spin-orbit entangled spin-1 coherent spin center: the manganese neutral acceptor in a III-V semiconductor}

\author{Julian Zanon}
\email{j.zanon@tue.nl}
\affiliation{Department of Applied Physics and Science Education, Eindhoven University of Technology, Eindhoven 5612 AZ, The Netherlands}

\author{Michael E. Flatt\'e}
\email{michaelflatte@quantumsci.net}
\affiliation{Department of Physics and Astronomy, University of Iowa, Iowa City, Iowa 52242, USA}
\affiliation{Department of Applied Physics and Science Education, Eindhoven University of Technology, Eindhoven 5612 AZ, The Netherlands}

\date{\today}

\begin{abstract}
% A Mn dopant in a semiconductor such as GaAs produces a highly-entangled, coherent spin ground state, beyond the capability of single-determinant electron-structure theories to fully capture. 
% The half-filled  $d$ states of the Mn form an $S=5/2$ core spin, which couples antiferromagnetically to a valence hole with orbit $L=1$ and spin $s=1/2$. The ground state triplet, $F=S+L+s=1$, is degenerate in the tetrahedral point group of the host semiconductor. We directly construct an analytic form for the ground-state triplet wavefunction, which exhibits surprising spin-charge correlation unlike that obtained in semiclassical calculations. Analytic expressions for spin-correlated circulating currents associated with the dopant yield remarkably large magnetic fringe fields of $\sim$1$\,\mu$T at distances of 8 nm from the manganese site. NV-diamond magnetometry would be able to distinguish the predicted fully quantum features from semiclassical ones while the dopant spin undergoes coherent precession.
A Mn dopant in a III-V semiconductor %such as GaAs 
produces a highly-entangled, coherent triplet ground state not fully captured by single-determinant theories of electron structure. We directly construct an analytic form for its ground-state wavefunction, finding surprising spin-charge correlations not revealed by semiclassical calculations. Spin-correlated circulating currents associated with the dopant yield remarkably large magnetic fringe fields of $\sim$1$\,\mu$T at distances of $\sim 10$~nm from  Mn in GaAs, potentially detectable by NV-diamond magnetometry while the dopant spin coherently precesses.
\end{abstract}

\maketitle
The ability to control matter at the atomic scale in a semiconductor  underlies %has been, undoubtedly, one of the major goals in the past decades. This profound interest has driven a huge development of
new technologies such as quantum computing \cite{Burkard2023, PhysRevA.57.120}, spintronics \cite{WU201061} and biological sensing \cite{Schirhagl2014, medintz2005quantum}. For these technologies fully localized (zero-dimensional) electronic states with internal degrees of freedom, such as those confined to quantum dots and near crystal impurities (often color centers), play pivotal roles. 
Examples of impurities in crystals which permit quantum coherent control of their spin-degree of freedom include P in Si \cite{Pla2012}, NV centers in diamond \cite{Schirhagl2014}, B in Si/SiO$_{2}$ \cite{van2018readout} and Mn in GaAs \cite{myers2008zero, PhysRevB.94.085308}.  At low concentrations, Mn in III-V semiconductors replaces a cation atom leaving a half filled 3d$^5$ shell, which provides a core spin equal to S$ = 5/2$ and binds a valence hole with antiparallel spin $J=3/2$ \cite{Schneider1987} to produce a highly spin-orbit entangled core-hole complex with total spin $F=1$. Extensive studies of the properties of individual Mn dopants using cross-sectional scanning tunneling microscopy, combined with   theoretical descriptions of the valence hole wave function, have effectively described many features of the measurements of a single Mn impurity at \cite{Kitchen2006} and below the surface \cite{Tang2004, Yakunin2004a}, under strain \cite{Yakunin2007, Jancu2008, Garleff2010}, interacting with nearby dopants \cite{Gupta,Kitchen2006,Yakunin2005} and for certain static external magnetic field directions \cite{Bozkurt2013}. In other hosts, like InAs \cite{Marczinowski2007}, GaP \cite{celebi2008} and InSb \cite{Mauger2015}, the spatial structure of Mn appears similar, as does the valence hole structure in silicon \cite{Acceptor_in_Si} (although viewed on the (001) surface instead of the (110) cleavage surface of III-V materials). Extending for approximately one nanometer, the Mn spatial structure is a consequence of a delocalization of the hole wave-function around the Mn core \cite{Schneider1987}, which could be potentially used within fast quantum gates \cite{A_M_Stoneham_2003} due to the strong coupling between the Mn impurity and external electric fields \cite{Tang2006}. However each of these treatments neglected the quantum nature of the Mn core spin and the full angular momentum of the core-hole complex. As a result these treatments are insufficient to describe the full quantum evolution of a Mn dopant in the bulk \cite{Nestoklon_2015, PhysRevLett.106.017202}.

Here we fully treat the core-hole spin entanglement by constructing a nearly analytic, full quantum theory of the complex and calculating the correlation between spin orientation and properties of the electronic state, including probability density, spin-correlated circulating orbital currents, and local magnetic fields. The $F=1$ character of the substitutional Mn core-hole spin complex is unsplit in the crystal field of the bulk III-V semiconductor, but has a huge response to local electric fields \cite{Tang2006}. Thus this quantum spin 1 provides an ideal complementary quantum entity to spin 1 dopants with very weak spin-orbit interaction (NV$^{-}$ in diamond, divacancies in SiC) \cite{PhysRevB.85.205203, koehl2011room, PhysRevLett.112.187601}, with a known long coherence time \cite{myers2008zero}, for exploring exceptionally rapid electric-field-driven spin manipulation. Our results are distinctly different from those obtained with semiclassical core-spin treatments that ignore entanglement of the core spin with the valence hole, suggesting that core-spin entanglement cannot be ignored for the proper theory of  Mn in GaAs. We note that this core-hole spin-orbit entanglement is  quenched by the Mn-Mn interactions in the higher concentration regime in which these Mn dopants drive ferromagnetism in Ga$_{1-x}$Mn$_x$As and other III-V dilute magnetic semiconductors \cite{Dietl2000, Dietl2002,WU201061, dietl2010ten}. 

We construct the full quantum $F=1$ basis from a hole spin J =$3/2$ oriented antiferromagnetically  with the Mn $3d^5$ core spin S = $5/2$, producing a total angular momentum spin $F$ = $S$ + $J$ = 1 (as described in Ref.~\onlinecite{Schneider1987}). The projections $m_{F} = -1, 0, 1$ along the z-direction for $F = 1$ form a new basis set $\{|1,m_{F}\rangle\}$  = $\{|1,+1\rangle$, $|1,0\rangle$ , $|1,-1\rangle\}$, where each element is expressed as

\begin{equation}
    |1,0\rangle = \sum_{m=-3/2}^{3/2}C_{m,-m}|m, -m\rangle, \ \label{eq: state_1_minus_1}
\end{equation}

\begin{equation}
    |1,\pm 1\rangle = \sum_{m = -1/2}^{5/2}C_{\pm m, \pm(1 - m)}|\pm m, \pm(1 - m)\rangle.\label{eq: state_1_plus_1}
\end{equation}
The right-hand side of Eqs.(\ref{eq: state_1_minus_1}) and  (\ref{eq: state_1_plus_1}) are proportional to a superposition of states with the form $|m_{S}, m_{J}\rangle$, where the $3d^5$ Mn core is described by $m_{S=5/2}$ and the hole with $m_{J=3/2}$. The Clebsch-Gordan coefficients $C_{m,n}$ follow in the Supplemental Material \cite{supplement_material_}.

The interdependence between core-hole spin of a single Mn and its spatial structure has already been proposed in other theoretical works \cite{Tang2005, Tang2006, Strandberg2009}, however they relied on a classical description of the core spin as a vector and neglected all the possible projections from a quantum-mechanical picture, i.e., $m_{S}$ =  -5/2, -3/2, -1/2, 1/2, 3/2, 5/2 for the Mn core spin S = 5/2. An \textit{ab initio} model for the Mn core spin was developed \cite{Mahani20142nd, Mahani20143rd} considering all the $d$ orbitals from the Mn 3$d^{5}$ shell. Such model gave the three-fold degeneracy for the acceptor state, expected for a state with a total angular momentum $S + J = 1$ in a cubic crystal, and the magnetic anisotropy energy had also a slight correction from a previous work \cite{Strandberg2009}. Nevertheless, as explained in \cite{Mahani20142nd}, the mean-field approximation in the \textit{ab initio} treatment also caused the Mn core spin to have only a particular projection, resembling the classical treatment where the different projections were not taken into account.

\begin{figure}[t!]
\begin{centering}
\includegraphics[scale=1]{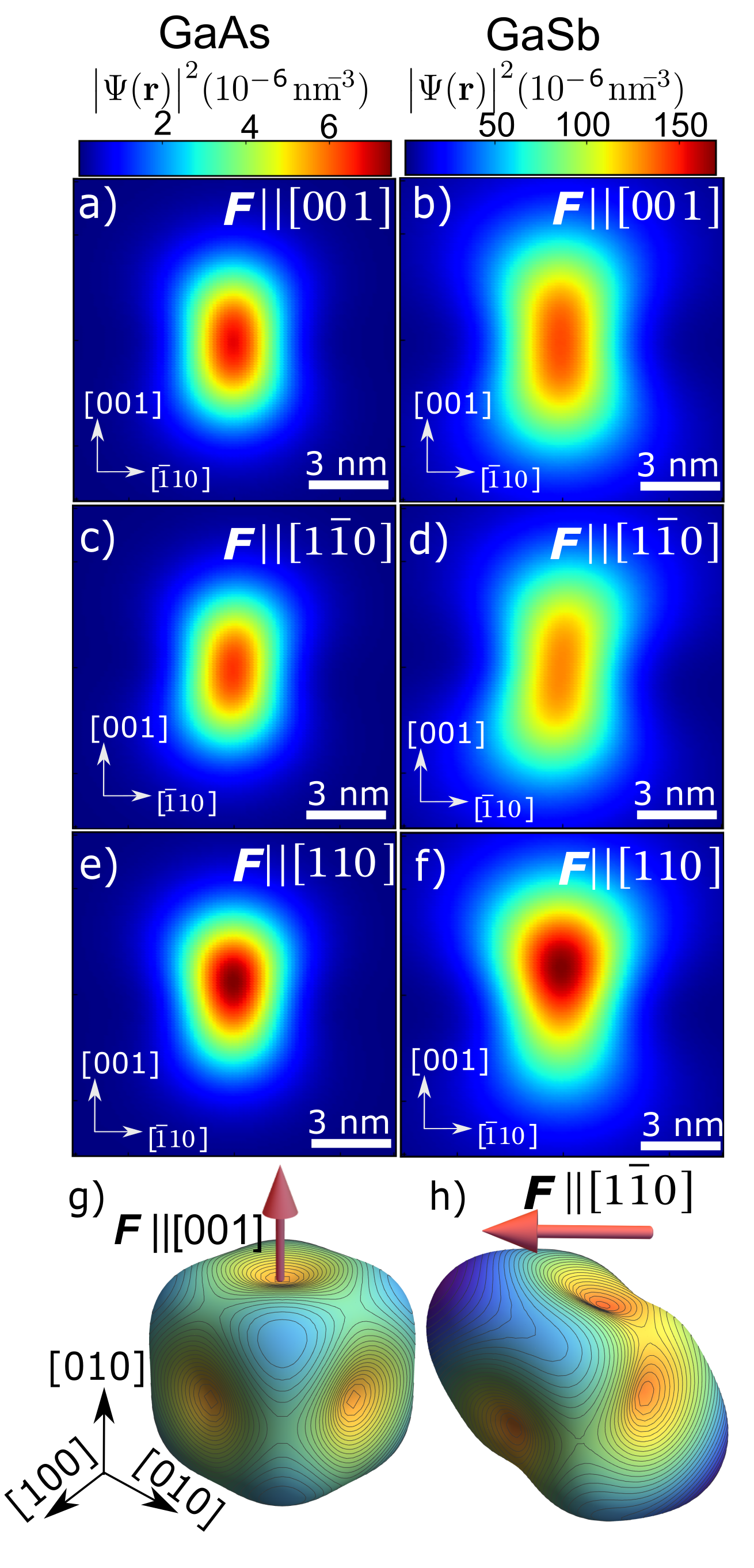}
\par\end{centering}
\raggedright{}\caption{\label{fig: Fig1_wave_functions}Probability density $|\Psi(\boldsymbol{r})|^{2}$ of a single Mn in GaAs and GaSb hosts within our fully quantum-mechanical description. All planes are 4.95 nm away from the impurity site, along the $[110]$ direction. From (a) to (f) each row has a fixed direction for the core-hole spin $\boldsymbol{F}$ and shows the differences between a Mn in GaAs or in GaSb. (g) and (h) show the isosurface of the probability density $|\Psi|^2$ (scaled by $e^{r/a_B^*}$) at $2.5\times 10^{-3}\,\text{nm}^{-3}$ for a Mn in GaSb  with $\boldsymbol{F}\parallel[001]$ and $\boldsymbol{F}\parallel[1\overline{1}0]$, respectively; $a_B^*=$0.77 nm (GaAs) and 2.27 nm (GaSb) (see Supporting Information). Binding energies: Mn in GaAs $E_{F=1}=113 $ meV \cite{PhysRevB.10.2501, madelung2004semiconductors}; Mn in GaSb $E_{F=1} =$ 18 meV \cite{madelung2004semiconductors}. Effective hole masses: GaAs $m_{lh} = 0.074$ and $m_{hh} = 0.559$; GaSb $m_{lh} = 0.041$ and $m_{hh} = 0.40$, from \cite{PhysRevB.8.2697,10.1063/1.1368156}.
}
\end{figure}

Compared to a classical description where the hole wave-function has the basis $\{|m_{J}\rangle\}$ as in \cite{Yakunin2004a, Yakunin2007} our basis set is given by the product of the Mn core and hole spin spaces, i.e., $\{|m_{S}, m_{J}\rangle\}$ = $\{|m_{S}\rangle\otimes| m_{J}\rangle\}$, see Eqs.(\ref{eq: state_1_minus_1}) and (\ref{eq: state_1_plus_1}). If $\hat{\mathcal{H}}_{0}$ is the Hamiltonian that describes the host with a single impurity, which creates an acceptor with a binding energy $E_{0}$, the core-hole complex is added through the exchange interaction between the Mn core spin $\boldsymbol{S}$ and the hole spin $\boldsymbol{J}$, i.e., $\hat{\mathcal{H}}=\hat{\mathcal{H}}_{0}+E_{exc}\hat{\boldsymbol{S}}\cdot\hat{\boldsymbol{J}}$ \cite{Monakhov2006, nestoklon2015fine,PhysRevB.107.174401}. $\hat{\mathcal{H}}$ is diagonal in the basis set $\{|1,m_{F}\rangle\}$ and the new binding energy $E_{F=1}$ is threefold degenerate $E_{F=1} \equiv$ $E_{m_F=\pm1,0}$ $=E_{0}-21\,E_{exc}/4$. The acceptor binding energy $E_{0}$ given by a Couloumb \cite{PhysRevB.8.2697} or a zero-like potential \cite{celebi2008, averkiev1994spin} is corrected by the exchange interaction energy $E_{ex}$. 

If we include a spin dependent effect into $\hat{\mathcal{H}}$, e.g., a zeeman term $g_{F}\mu_{B}$$\boldsymbol{H}\cdot\boldsymbol{F}$ where $\boldsymbol{H}$ is the external magnetic field, $\boldsymbol{F}$ the core-hole spin, $\mu_{B}$ the Bohr magneton and $g_{F} = 2.77$ given from \cite{Schneider1987}, a general eigenstate for the system has the form 
\begin{equation}
\big|\Psi\rangle=\text{a}_{-1}\big|1,-1\rangle+\text{a}_{0}\big|1,0\rangle+\text{a}_{+1}\big|1,+1\rangle,
\label{eq: total_psi_F_1}
\end{equation}
with $\{\text{a}_{m_{F}}\}$ being complex numbers. To verify how the hole probability density (i.e., the Mn spatial structure) is affected for a given direction of $\boldsymbol{F}$,
\begin{comment}
\textcolor{red}{we consider that $|\boldsymbol{H}|$ is small enough such that the zeeman splitting for $E_{F}$ is comparable to the thermal fluctuation energy}. In this approximation,
\end{comment} 
we keep the binding energy $E_{F}$ equal to its bulk value and the coefficients $\{\text{a}_{m_{F}}\}$ are chosen such that $\langle\Psi\big|\boldsymbol{F}\big|\Psi\rangle$$\parallel$ $\boldsymbol{\hat{n}}$ for a given spatial direction $\boldsymbol{\hat{n}}$, see Tab.{\color{blue} I} in the Supplementary Material \cite{supplement_material_}. We find the probability density  $\big|\Psi\,(\boldsymbol{r})\big|^{2}$ = $\langle \Psi|\boldsymbol{r}\rangle\langle \boldsymbol{r}|\Psi\rangle$ considering the analytical hole wave-functions $\{\langle \boldsymbol{r}|m_{J}\rangle\}$ obtained within the effective mass approximation, and treating the impurity as a zero-like potential \cite{averkiev1994spin, Yakunin2004a}. This procedure, as presented in the Supplemental Material \cite{supplement_material_}, ignores the wave-function spatial dependency of the Mn core (because of its localized nature) and allows us to have the following analytical form for the hole probability density.

Eq.(\ref{eq: probability_density}) depends on the distance from the impurity site $r$, the azimuthal angle $\theta$, the polar angle $\varphi$ and the envelope functions $R_{0}(r)$ and $R_{2}(r)$. Besides the exponential dependence on the distance from the manganese site, $R_{0}(r)$ and $R_{2}(r)$ depend on the acceptor binding energy $E_{F=1}$ and on the light- and heavy-hole effective masses (i.e., $m_{lh}$ and $m_{hh}$, respectively), see Supplemental Material \cite{supplement_material_}. Using Eq.(\ref{eq: probability_density}), the probability density for a single Mn in GaAs and in GaSb with different core-hole spin directions is presented in Fig.\ref{fig: Fig1_wave_functions}. We present the calculations at the (110) plane, since it is a normal cleavage plane used back in X-STM images \cite{Yakunin2004a}.

\begin{widetext}
\begin{align}
\big|\Psi(\boldsymbol{r})\big|^{2} & =  \frac{\big|R_{0}(r)\big|^{2}}{4\pi}+\frac{5\,\big|R_{2}(r)\big|^{2}}{4\pi}\,\big[\text{cos}^{2}(\theta)\text{sin}^{2}(\theta)+\text{sin}^{2}(\varphi)\,\text{cos}^{2}(\varphi)\,\text{sin}^{4}(\theta)\big]+ \big\{\big(\text{a}_{+1}^{*}\text{a}_{-1}-\text{a}_{-1}^{*}\text{a}_{+1}\big)\frac{i\,\sin^{2}(\theta)\,\sin(2\varphi)}{3\sqrt{2}}\nonumber \\
& -\big(\text{a}_{0}^{*}\text{a}_{+1}+\text{a}_{-1}^{*}\text{a}_{0}\big)e^{i\varphi}\sin(\theta)\cos(\theta) -\big(\text{a}_{+1}^{*}\text{a}_{0}+\text{a}_{0}^{*}\text{a}_{-1}\big)e^{-i\varphi}\sin(\theta)\cos(\theta)\big\} \frac{3\sqrt{30}}{40\pi}R_{0}(r)R_{2}(r).
\label{eq: probability_density}
\end{align}
\end{widetext}

Figs.\ref{fig: Fig1_wave_functions}(a) and (b) show that for $\boldsymbol{F}\parallel[001]$, the probability density has a `bow tie' like shape. This symmetry is a result only from the first two terms on the right-hand side of Eq.(\ref{eq: probability_density}), because for $\boldsymbol{F}\parallel[001]$ we have $a_{+1}$=1 while $a_{0}=a_{-1}$=0. The `bow tie' is similar to what was reported using X-STM and the effective mass model in \cite{Yakunin2004a}, and such shape for the probability density at the (110) plane reflects the cubic symmetry that states close to edge of the valence band in III-V hosts develop, see also Fig.\ref{fig: Fig1_wave_functions}(g). Besides that, as already discussed in \cite{mahadevan2004trends} the spatial structure of a Mn in GaAs is more localized than in GaSb, because of its larger binding energy and smaller effective mass ratio ($m_{lh}$/$m_{hh}$). 

In contrast to the cubic symmetry, the probability density exhibits a distortion either when $\boldsymbol{F}\parallel[110]$ or $\boldsymbol{F}\parallel[1\overline{1}0]$ as presented in Figs.\ref{fig: Fig1_wave_functions}(c)-(f). In addition to that, the probability density deforms along the direction of the spin, which is surprisingly different from what was reported in Refs.\cite{Tang2005, Bozkurt2013}, where a classical description for the Mn core spin shows a distortion for the local density of states at the plane perpendicular to the direction of the spin. In other words, at the (110) plane in Figs.\ref{fig: Fig1_wave_functions}(a)-(f)  the probability density for $\boldsymbol{F}\parallel[110]$ deforms qualitatively similar to what was calculated in \cite{Bozkurt2013} for a classical core spin $\boldsymbol{S}$ parallel either to [001] or to $[1\overline{1}0]$. Nonetheless, for $\boldsymbol{F}\parallel[1\overline{1}0]$ our result show a probability density slightly rotated clock-wise, quite different from what was reported in \cite{Bozkurt2013} for $\boldsymbol{S}\parallel[110]$.

%\begin{figure}[b]
%\begin{centering}
%\includegraphics[scale=0.83]{figures_paper/Fig1_wave_functions.png}
%\par\end{centering}
%\raggedright{}\caption{\label{fig: Fig1_wave_functions}Probability density $|\Psi(\boldsymbol{r})|^{2}$ of a single Mn in GaAs and GaSb hosts within our fully quantum-mechanical description. All planes are 4.95 nm away from the impurity site, along the $[110]$ direction. From (a) to (f) each row has a fixed direction for the core-hole spin $\boldsymbol{F}$ and shows the differences between a Mn in GaAs or in GaSb. (g) and (f) present a 3D view of the hole probability density $3.5$ nm away from the Mn site in the GaSb host, with $\boldsymbol{F}\parallel[001]$ and $\boldsymbol{F}\parallel[1\overline{1}0]$, respectively. Binding energies: Mn in GaAs $E_{F=1}=113 $ meV \cite{PhysRevB.10.2501, madelung2004semiconductors}; Mn in GaSb $E_{F=1} =$ 18 meV \cite{madelung2004semiconductors}. Effective hole masses: GaAs $m_{lh} = 0.074$ and $m_{hh} = 0.559$; GaSb $m_{lh} = 0.041$ and $m_{hh} = 0.40$, from \cite{PhysRevB.8.2697,10.1063/1.1368156}.
%}
%\end{figure}

\begin{figure}
\begin{centering}
~\includegraphics[scale=1]{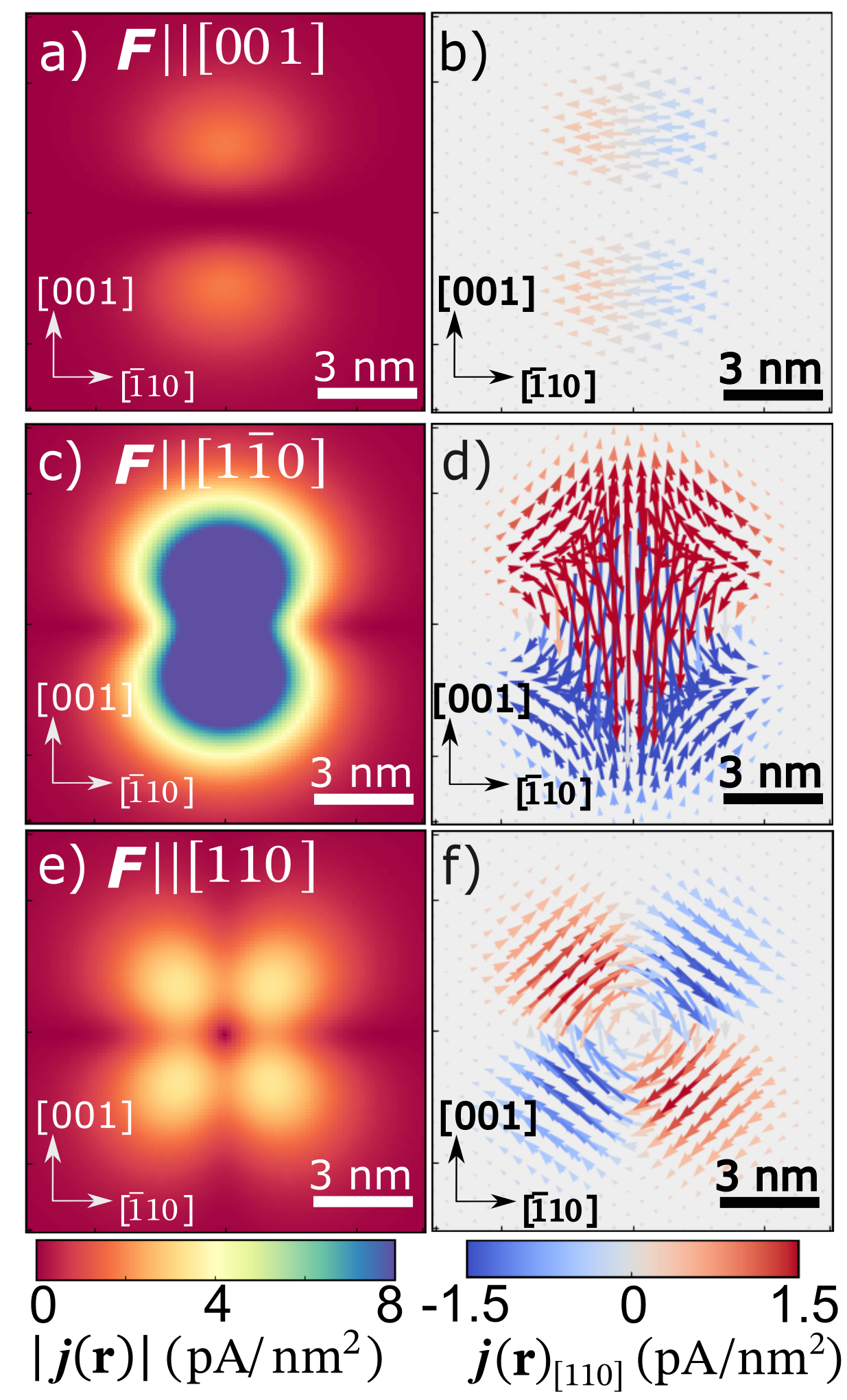}
\par\end{centering}
\centering{}\caption{\label{fig: Fig2_current_density} Current density $\boldsymbol{j}(\boldsymbol{r})$ for different core-hole spin $\boldsymbol{F}$ directions, for a single Mn in GaAs. All planes are 4.95 nm away from the impurity site, along the [110] direction. (a), (c) and (e) show the absolute values for the current density, while (b), (d) and (f) are the field vectors. The right-hand panels have  arrow directions defined by $\boldsymbol{j}(\boldsymbol{r}) \parallel [001]$ and $\boldsymbol{j}(\boldsymbol{r}) \parallel [\overline{1}10]$;  arrow colors show the component $\boldsymbol{j}(\boldsymbol{r})_{[110]}$ $ \equiv \boldsymbol{j}(\boldsymbol{r})\parallel [110]$ and the length of the arrow is given by $\big|\boldsymbol{j}(\boldsymbol{r})\big|$, where $\big|\boldsymbol{j}(\boldsymbol{r})\big|$ is shown in the left-hand panels.}
\end{figure}

Such strong spin-orbit interdependency observed in the probability density, Fig.\ref{fig: Fig2_current_density}, could generate a change in the circulating current as already calculated for other defect systems in  \cite{PhysRevLett.112.187201,circulating_current_Adonai}. To calculate the circulating current, we use the general expression \cite{landau1958course} 

\begin{equation}
\begin{aligned}
    \boldsymbol{j}(\boldsymbol{r}) &= 2\,\mu_{B}\,\text{Im}\big\{\Psi^{\dagger}(\boldsymbol{r})\boldsymbol{\nabla}\Psi(\boldsymbol{r})\big\} - 2\mu_{B}\,\boldsymbol{\mathcal{A}}\big|\Psi(\boldsymbol{r})\big|^2 \nonumber \\ 
    & + g_{F}\mu_{B}\boldsymbol{\nabla}\times\big(\frac{1}{\hbar}\Psi^{\dagger}(\boldsymbol{r})\,\boldsymbol{F}\,\Psi(\boldsymbol{r})\big),
    \label{eq: current_density}
\end{aligned}
\end{equation}
where the factor $g_{F}=2.77$ was measured by \cite{Schneider1987} and $\mu_{B}$=$9.274\times10^{-6}$$\rm{A}\cdot\rm{nm}^2$ is the Bohr magneton given in Amperes ($\rm{A}$) nanometers (nm) squared.  

The first term in Eq.(\ref{eq: current_density}) is associated with the momentum operator, while the second is a result of the minimal coupling (e.g., due to an external magnetic field) and the third explicitly depends on the spin $\boldsymbol{F}$. The second and third terms in Eq.(\ref{eq: current_density}) will provide contributions to the magnetic fringe fields (later calculated) that are at least two and four orders of magnitude smaller than the first term, respectively (see Supplemental Material \cite{supplement_material_}).Therefore, only the momentum operator contribution is considered. 

Fig.\ref{fig: Fig2_current_density} presents the circulating current $\boldsymbol{j}(\boldsymbol{r})$ around a single Mn in GaAs, with its core-hole spin pointing to three different directions. Looking at Figs.\ref{fig: Fig2_current_density}(b), (d) and (f) the complex circulating pattern of $\boldsymbol{j}(\boldsymbol{r})$ occurs in a plane perpendicular to the core-spin direction, a similar effect was calculated for a defect in a InAs 2DEG \cite{circulating_current_Adonai}. In our theory, $\boldsymbol{j}(\boldsymbol{r})$ is a consequence of the terms that go along with the d-like envelope function $R_{2}(r)$, in the simplest case it does not only give the `bow tie' symmetry in Figs.\ref{fig: Fig1_wave_functions}(a) and (b) but also causes $\boldsymbol{j}(\boldsymbol{r})\neq\boldsymbol{0}$ in Fig.\ref{fig: Fig2_current_density}(a). When $\boldsymbol{F}\parallel[1\overline{1}0]$ or $\boldsymbol{F}\parallel[110]$, angular dependent terms contribute to a circulating current with a `bow tie' or a d-like symmetry in Figs.\ref{fig: Fig2_current_density}(c) and (e), respectively. The strongest intensity for $\boldsymbol{j}(\boldsymbol{r})$ at the (110) plane occurs for $\boldsymbol{F}\parallel[1\overline{1}0]$, with a maximum value that goes up to 15 pA/nm$^{2}$; the color scale has a maximum of $8$\,pA/nm$^{2}$ to facilitate the visualization of small current values. In the Supplementary Material \cite{supplement_material_}, we also verify that a single Mn in InSb (GaSb) generates a circulating current that goes up to  $400$ pA/nm$^{2}$ ($300$ pA/nm$^{2}$); this is two orders of magnitude larger than a Mn in GaAs. This large difference is attributed to the binding energy and effective mass ratio $\,m_{lh}/m_{hh}\,$ which in InSb is larger than in GaSb and for GaAs those parameters are the largest.

\begin{figure}[h!]
\begin{centering}
~\includegraphics[scale=1]{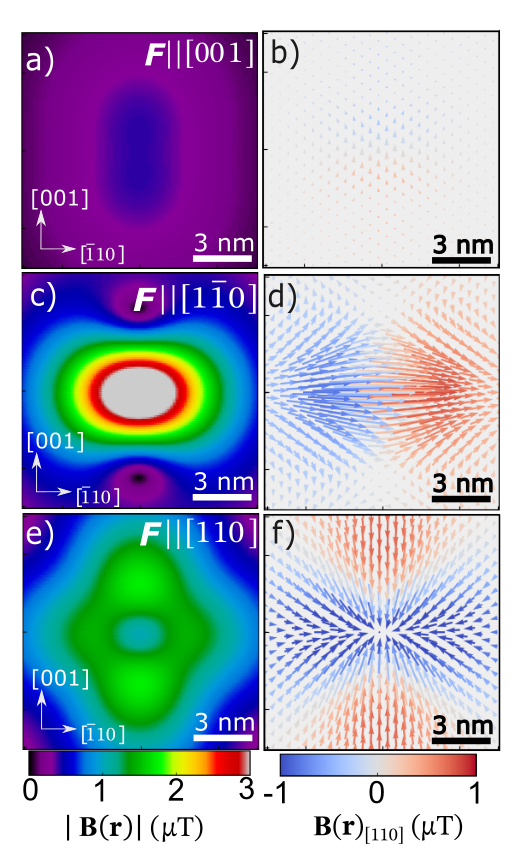}
\par\end{centering}
\centering{}\caption{\label{fig: Fig3_fringe_fields_plane}Magnetic fringe fields $\boldsymbol{B}(\boldsymbol{r})$ at a plane along the [110] direction and 4.95 nm away from the Mn site in GaAs. Each row corresponds to a different core-hole spin direction $\boldsymbol{F}$. (a), (c) and (e) show the absolute values for the magnetic field, while (b), (d) and (f) are the field vectors. The right-hand panels have  arrow directions defined by $\boldsymbol{B}(\boldsymbol{r}) \parallel [001]$ and $\boldsymbol{B}(\boldsymbol{r}) \parallel [\overline{1}10]$;  arrow colors show the component $\boldsymbol{B}(\boldsymbol{r})_{[110]}$ $ \equiv \boldsymbol{B}(\boldsymbol{r})\parallel [110]$ and the length of the arrow is $\big|\boldsymbol{B}(\boldsymbol{r})\big|$, which is shown in the left-hand panels.}
\end{figure}

The magnetic fringe field related to these circulating currents are evaluated numerically with the Biot-Savart law \cite{jackson1998classical}

\begin{equation}
    \boldsymbol{B}(\boldsymbol{r}) = \frac{\mu_{0}}{4\pi}\int\,d^{3}\boldsymbol{r}'\,\frac{\boldsymbol{j}(\boldsymbol{r}')\times(\boldsymbol{r}-\boldsymbol{r}')}{|\boldsymbol{r}-\boldsymbol{r}'|^{3}},
    \label{eq: magnetic_field_biot-savart}
\end{equation}
with $\mu_{0} = 4\pi\times10^{2}\,\rm{H/nm}$. 

\begin{figure}
%\begin{centering}
\includegraphics[scale=1]{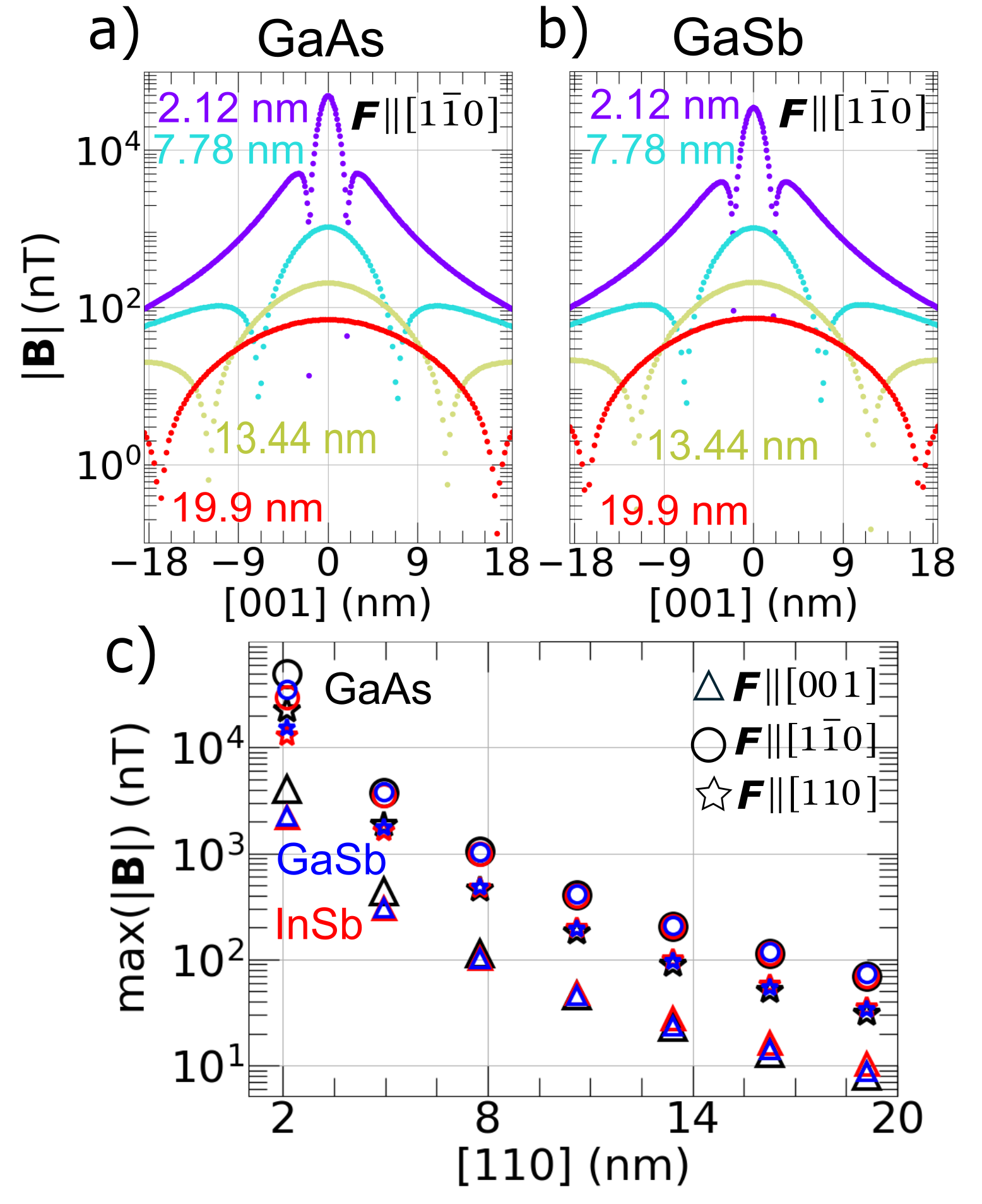}
%\par\end{centering}
\caption{\label{fig: Fig4_fringe_field_comparison}Magnetic field profiles along the [001] direction produced by a core-hole spin $\boldsymbol{F}\parallel[1\overline{1}0]$ of a single Mn in  (a) GaAs and  (b) GaSb. Each line's  color corresponds to the indicated (110) plane distance (in nm) from the Mn site. (c)  maximum value for the magnetic field max$(|\boldsymbol{B}|)$ at several (110) planes, produced by a single Mn in three different hosts (GaAs, InSb and GaSb)  with the core-hole spin $\boldsymbol{F}$ pointing in three different  directions. Parameters for InSb: binding energy $E_{F=1}=9$ meV \cite{Mauger2015, teubert2009influence} and  effective masses for the light-hole $m_{lh}$=0.014 and heavy-hole $m_{hh}=$0.44\cite{PhysRevB.8.2697,10.1063/1.1368156}.}
\end{figure}

Fig.~\ref{fig: Fig3_fringe_fields_plane}a) shows that a single Mn in GaAs with its core-hole spin $\boldsymbol{F}\parallel[001]$ produces a magnetic field gradient reaching up to $0.5 \,\mu\rm{T}$, on a (110) plane located 4.95 nm away from its site. When its core-hole spin changes to 
 $\boldsymbol{F}\parallel[110]$ the field increases to $1.5 \,\mu\rm{T}$ and its structure changes significantly, as shown in Fig.~\ref{fig: Fig3_fringe_fields_plane}e). For $\boldsymbol{F}\parallel[1\overline{1}0]$ the field strength increases substantially, reaching $3.5 \,\mu\rm{T}$, and its shape acquires a `bow tie' symmetry along the $[1\overline{1}0]$ direction, Fig.~\ref{fig: Fig3_fringe_fields_plane}c). In Figs.~\ref{fig: Fig3_fringe_fields_plane}(b), (d) and (f) the magnetic field presents a complex nanoscale orientation on the (110) plane.   

The magnetic field strength depends on the distance between the Mn site and the (110) plane. For GaAs in Fig.~\ref{fig: Fig4_fringe_field_comparison}(a) the magnetic field profile along the [001] direction shows a decreasing in strength (and a broadening of its feature) from $\sim$$0.5\times10^{5}\,$nT, at the (110) plane 2.12 nm away from the Mn site, to $\sim$$10^{2}$ nT at the (110) plane 19.9 nm away from the Mn site; a similar result is obtained for a Mn in GaSb see Fig.~\ref{fig: Fig4_fringe_field_comparison}b). Fig.~\ref{fig: Fig4_fringe_field_comparison}c) shows that the magnetic field produced by a core-spin $\boldsymbol{F}\parallel[1\overline{1}0]$, at various distances along the [110] direction, is predominantly larger than for $\boldsymbol{F}\parallel[001]$ or $\boldsymbol{F}\parallel[110]$. At a given (110) plane and a spin-core orientation, the magnetic field from a Mn in GaAs is slightly larger than in GaSb or InSb. This is likely due to the current density  being more localized in GaAs, which causes the integral in Eq.(\ref{eq: magnetic_field_biot-savart}) to converge faster (as $\boldsymbol{r}'\rightarrow \infty$) than in GaSb or in InSb.

Detecting such magnetic fringe fields could be possible using magnetometers based on NV-centers \cite{Degen_2008, doi:10.1073/pnas.0812068106}. For example, a magnetic field sensitivity of $\sim 50$ nT/$\sqrt{\rm{Hz}}$ was reached in \cite{10.1063/1.4952953}, which enabled the detection of stray fields of a single Ni nanorod with resolutions of a few tens of nanometers. Reducing the NV-to-sample distances would likely improve further the spatial resolution, a recent work \cite{xu2025minimizing} demonstrated NV-to-sample distances of 7.9 $\pm$ 0.4 nm enabling spatial resolution of few nanometers for magnetic fields of a few hundreds of $\mu\,\rm{T}$. The NV magnetometer could be used over a (110) plane of a GaAs quantum well, where a few atoms of Mn could have its core-hole spin oriented using a zero-field optical manipulation, as done before \cite{myers2008zero}. Such detection would be an indirect access to the core-hole spin orientation and, since a single Mn has a large hole wave-function, a few Mn impurities hosted in III-V semiconductors could possibly be used as fast quantum-gates due to their coupling with external electric fields \cite{Tang2006}.

%\section{Acknowledgments}
J.~Z.  was financially supported by Marie Sklodowska-Curie  Grant No. 956548.

%\appendix%

\bibliography{library}

\end{document}